\documentclass[aps,prb,reprint,superscriptaddress]{revtex4-1}

\usepackage{soul}
\usepackage{amsmath}
\usepackage{amsfonts}
\usepackage{amssymb}
\usepackage[english]{babel}
\usepackage[amssymb]{SIunits}
\usepackage{graphicx}
\usepackage{dcolumn}
\usepackage{bm}
\usepackage{color}

\begin{document}

\title{Manipulating quantum Hall edge channels in graphene through Scanning Gate Microscopy}

\author{Lennart Bours}
\affiliation{NEST, Istituto Nanoscienze--CNR and Scuola Normale Superiore, Piazza San Silvestro 12, 56127 Pisa, Italy}

\author{Stefano Guiducci}
\affiliation{NEST, Istituto Nanoscienze--CNR and Scuola Normale Superiore, Piazza San Silvestro 12, 56127 Pisa, Italy}

\author{Alina Mre\'nca--Kolasi\'nska}
\affiliation{AGH University of Science and Technology, Faculty of Physics and Applied Computer Science, al.~Mickiewicza 30, 30--059 Krak\'ow, Poland}

\author{Bart\l{}omiej Szafran}
\affiliation{AGH University of Science and Technology, Faculty of Physics and Applied Computer Science, al.~Mickiewicza 30, 30--059 Krak\'ow, Poland}

\author{Jan C Maan}
\affiliation{Radboud University Nijmegen, High Field Magnet Laboratory, Toernooiveld 7, 6525 ED Nijmegen, The Netherlands}

\author{Stefan Heun}%
 \email{stefan.heun@nano.cnr.it}
\affiliation{NEST, Istituto Nanoscienze--CNR and Scuola Normale Superiore, Piazza San Silvestro 12, 56127 Pisa, Italy}

\date{\today}

\begin{abstract}
We show evidence of the backscattering of quantum Hall edge channels in a narrow graphene Hall bar, induced by the gating effect of the conducting tip of a Scanning Gate Microscope, which we can position with nanometer precision. We show full control over the edge channels and are able, due to the spatial variation of the tip potential, to separate co-propagating edge channels in the Hall bar, creating junctions between regions of different charge carrier density, that have not been observed in devices based on top-- or split--gates. The solution of the corresponding quantum scattering problem is presented to substantiate these results, and possible follow--up experiments are discussed.
\end{abstract}

\maketitle

\section{Introduction}

The quantum Hall effect has been studied extensively since its discovery in 1980,\cite{Klitzing1980,VonKlitzing1986} and is one of the few quantum mechanical phenomena which are appreciable at a macroscopic level. It arises in two--dimensional systems as a consequence of gauge invariance,\cite{Laughlin1981} and its most striking feature is the exact quantization of the transverse Hall resistance, which is not sensitive to a moderate amount of disorder.

Due to graphene's Dirac--like dispersion, the cyclotron frequency $\omega_c$, and by extension the Landau level energy, scales with $\sqrt{B}$ as opposed to $B$ in classical Hall physics:\cite{CastroNeto2009,Sarma2010} $E_\pm \left( N \right) = \pm \hbar \omega_c \sqrt{N}$, where $N = 0, 1, 2, \ldots$ is a positive integer and $\omega_c = v_F \sqrt{2eB / \hbar }$.

Every Landau level is fourfold degenerate; two times for spin and two times for valley. The important exception is the anomalous zero energy Landau level, a consequence of graphene's non--trivial Berry phase of $\pi$ at the Dirac point, which is shared by electrons and holes, providing two extra channels to either regime. This gives the ``half--integer'' sequence of filling factors
\begin{equation}
\nu = \pm g_s g_v \left( N + \frac{1}{2} \right), \label{eq1}
\end{equation}
with $N = 0, 1, 2, \ldots$, and $g_s = g_v = 2$ the spin degeneracy and valley degeneracy in graphene, respectively.

Graphene's special bipolar nature further enriches its quantum Hall physics.\cite{Zhang2005,Novoselov2006,Novoselov2007} Through local electrostatic gating, junctions of opposite polarity can be created. Edge channels corresponding to regions of different charge carrier polarity\cite{Williams2007,Abanin2007,Ozyilmaz2007,Velasco2009,Ki2009} will flow in opposite directions, i.e.~electrons will flow clockwise, while holes will propagate counter--clockwise (or vice versa). At the interface between two regions of opposite polarity, edge channels of both regions will co--propagate, and their chemical potential will equilibrate, assuming the interface is sufficiently long.\cite{Alphenaar1990} Gating therefore allows one to manipulate quantum Hall edge channels. Under suitable parameters, the branching, equilibration, and complete backscattering of edge channels can be induced. 

These effects have been studied in graphene Hall bars with a central top--gate patterned on them.\cite{Abanin2007,Ozyilmaz2007,Ki2009} Depending on the filling factors on both sides of the junction, current can be backscattered in several ways. One can distinguish between three scenarios: partial, indirect, and complete backscattering, and calculate the expected value of $R_{xx}$ using the Landauer--B\"uttiker formalism.\cite{Buttiker1988}

If the central region has the same polarity and a smaller filling factor than the bulk $\left( \left| \nu' \right| < \left| \nu \right| \right)$, which gives a $p/p'/p$ or $n/n'/n$ junction, some edge channels will be backscattered, while others are transmitted; this is the partial backscattering regime, and the longitudinal resistance is\cite{Ki2009}
\begin{equation}
R_{xx} = \frac{h}{e^2} \frac{\left| \nu \right| - \left| \nu' \right|}{\left| \nu \right| \left| \nu' \right|}. \label{eq2}
\end{equation}

If the central region has the same polarity as the bulk, but a higher filling factor $\left( \left| \nu' \right| > \left| \nu \right| \right)$, current will be transported from one edge to another by edge channels which are localized in the central region, and equilibrate on both sides of the device. This indirect backscattering manifests itself as\cite{Ki2009}
\begin{equation}
R_{xx} = \frac{h}{e^2} \frac{\left| \nu' \right| - \left| \nu \right|}{\left| \nu \right| \left| \nu' \right|}. \label{eq3}
\end{equation}

When there are regions of different polarity, i.e.~a $p/n/p$ or $n/p/n$ junction, the backscattering will be direct and complete. Due to equilibration between the $n$-- and $p$--type channels at the interface, current will still be transmitted. Now\cite{Ki2009}
\begin{equation}
R_{xx} = \frac{h}{e^2} \frac{\left| \nu \right| + \left| \nu' \right|}{\left| \nu \right| \left| \nu' \right|}. \label{eq4}
\end{equation}

Hence, interactions between edge channels typically manifest themselves as fractional values of the von Klitzing constant $R_{K} = \frac{h}{e^2}$. By controlling this process, one can explore the interaction between channels and investigate their microscopic structure. Studying these processes provides an opportunity to investigate the elusive and long debated microscopic structure of the edge channels.

\begin{figure}[b]
	\includegraphics[width=0.8\columnwidth]{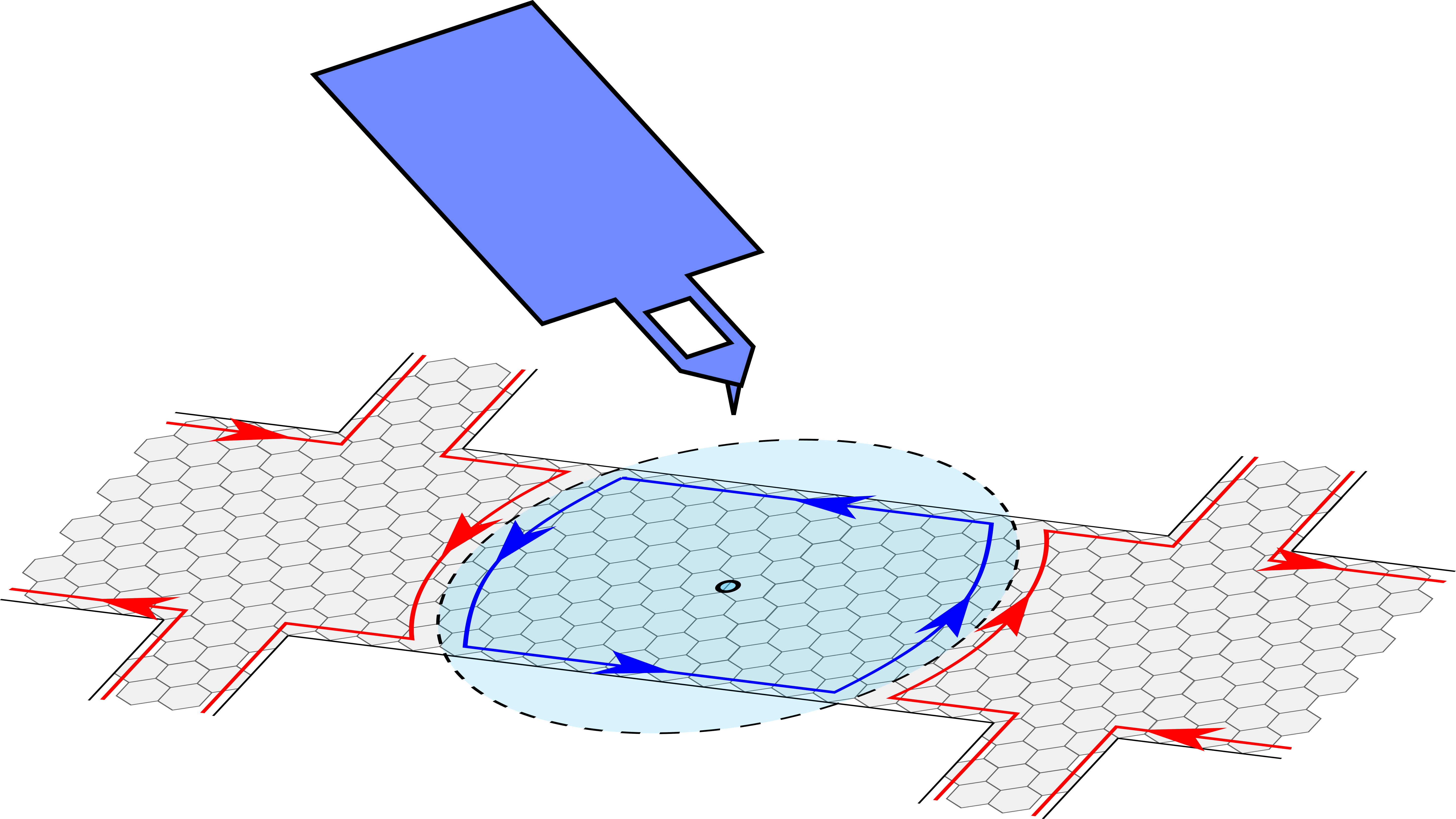}
	\caption{\label{fig:sgm_sketch} A schematic representation of the SGM set--up. By applying a voltage to the metallic tip, we can locally gate a region of choice.}
\end{figure}

In this article we demonstrate that we are able to \emph{locally} gate a region of choice of a graphene Hall bar, by applying a potential to the tip of a Scanning Gate Microscope (SGM), see Fig.~\ref{fig:sgm_sketch}. The SGM set--up consists of an Atomic Force Microscope (AFM) with a metallic (tungsten) tip. It is possible to gate a region of the sample located underneath the tip, by applying a voltage to the tip.\cite{Mrenca-Kolasinska} The SGM set-up allows us to explore the quantum Hall edge channel physics through a different and rarely used, non-uniform gating potential and provides full control over the edge channels, which can be made to interact, equilibrate, and/or backscatter as one sees fit.\cite{Paradiso2010, Paradiso2012, Paradiso2012a} As a consequence of the spatial variation of the tip potential, we are able to spatially separate co-propagating channels inside the uniform Hall bar, which leads to new, intricate junctions that have not been reported before. Finally, we demonstrate the ability to accurately simulate our experiments through tight binding simulations.

The paper is organized in the following way: In section~\ref{sec:methods} we present our Methods, which features a description of the SGM set--up and of the device. In section~\ref{sec:results} we present the experimental results. Furthermore, in section~\ref{sec:simul} we present quantum transport calculations, which add further support to our observations. In section~\ref{sec:discuss} the results are discussed, and in section~\ref{sec:summary} we shift our gaze forward and discuss how the work presented here is an important stepping stone for further research, which takes full advantage of the unique possibilities that Scanning Gate Microscopy offers.

\section{\label{sec:methods}Methods}

All measurements were performed at \unit{4.2}{\kelvin} and in a magnetic field of \unit{8}{\tesla} in the SGM cryostat. The electrical resistances were determined using a constant AC current of $\unit{100}{\nano\ampere}$, using lock--in amplifiers in a four probe configuration. DC voltage sources are used to bias the backgate and the tip.

The SGM system consists of a modified commercial Attocube tuning fork based AFM system. Tungsten tips are prepared by chemical etching and glued to the tuning fork. The AFM operates in non--contact mode and detects the reduction of the oscillation amplitude that arises due to shear forces between the tip and the sample. A stack of piezo elements allows for both fine movement, within a range of \unit{30}{\micro\meter}, and course movement in a range of \unit{5}{\milli\meter}. The entire system is designed to be isolated from vibrations and noise.

A reference AFM scan of the central part of the device is made before every measurement to compensate for possible drift. Using this scan we can position and move the tip to any desired position with great precision. Once the tip is in position, the Hall resistances $R_{xx}$ and $R_{xy}$ are measured to determine the effect of the tip as a function of tip position and bias.

\begin{figure}[t]
	\includegraphics[width=0.9\columnwidth]{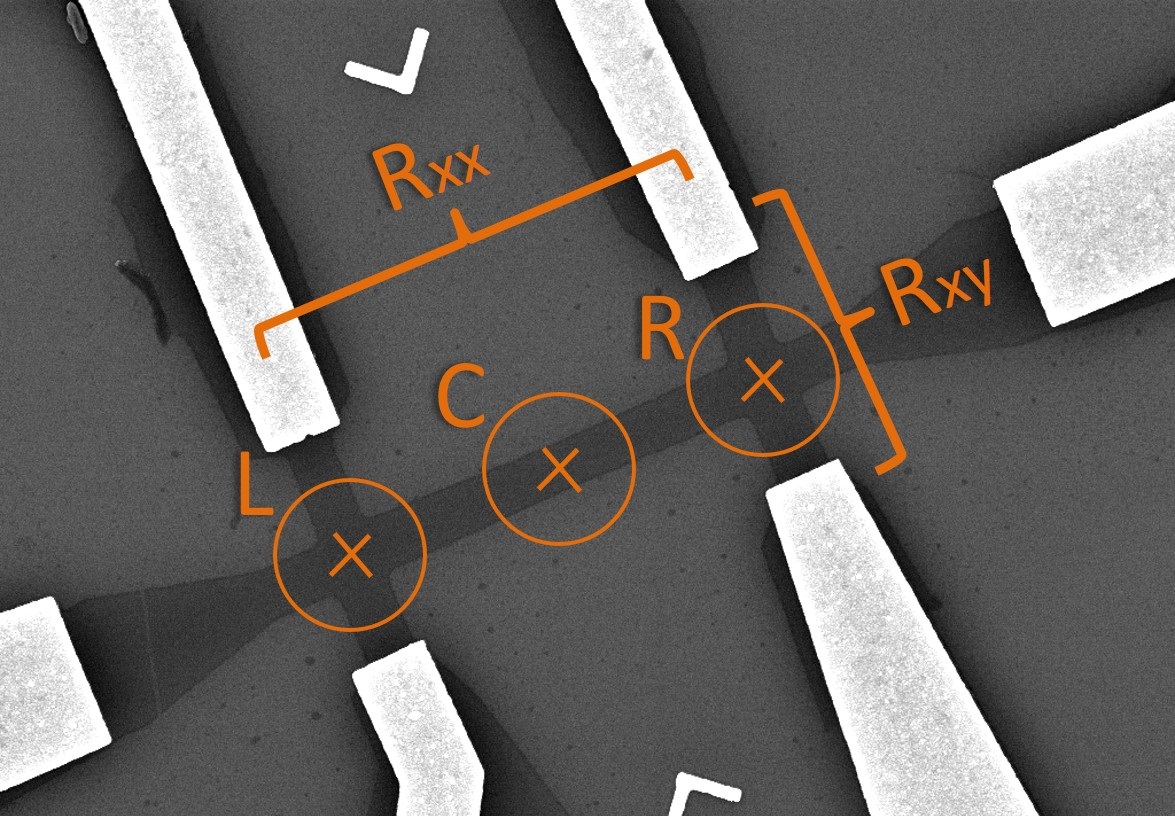}
	\caption{\label{fig:sem}A scanning electron microscope image of a device of the same design as the one described in this paper. The distance between the side contacts is 6 $\mu$m, while the width of the Hall bar is about \unit{800}{\nano\meter}. The probes used to measure $R_{xx}$ and $R_{xy}$ are indicated, as are the tip positions: left (L), centre (C), and right (R).}
\end{figure}

The device (see Fig.~\ref{fig:sem}) is fabricated from single layer graphene,\cite{Ferrari2006,Blake2007} exfoliated on a PVA/PMMA substrate and transferred to a Si/SiO$_2$ substrate with a \unit{300}{\nano\meter} thick  SiO$_2$ layer, where the highly doped Si acts as a backgate. Using standard nano--fabrication technologies such as electron beam lithography and reactive ion etching, the graphene is patterned into a Hall bar with length $L \approx$ \unit{6}{\micro\meter} and width $W \approx$ \unit{800}{\nano\meter}. These dimensions have been selected such that the Hall bar is sufficiently wide to allow for the unobstructed flow of edge channels in absence of any intervention,\cite{Han2007,Duerr2012} while being sufficiently narrow such that the channels can be made to interact through the gating effect of the SGM. To reduce the invasiveness and the screening effect of the metal contacts (Cr/Au \unit{10/60}{\nano\meter} thick),\cite{Huard2008,Pi2009,Liu2013} the Hall bar is connected through \unit{1}{\micro\meter} long graphene leads. 

\begin{figure}[t]
	\includegraphics[width=0.7\columnwidth]{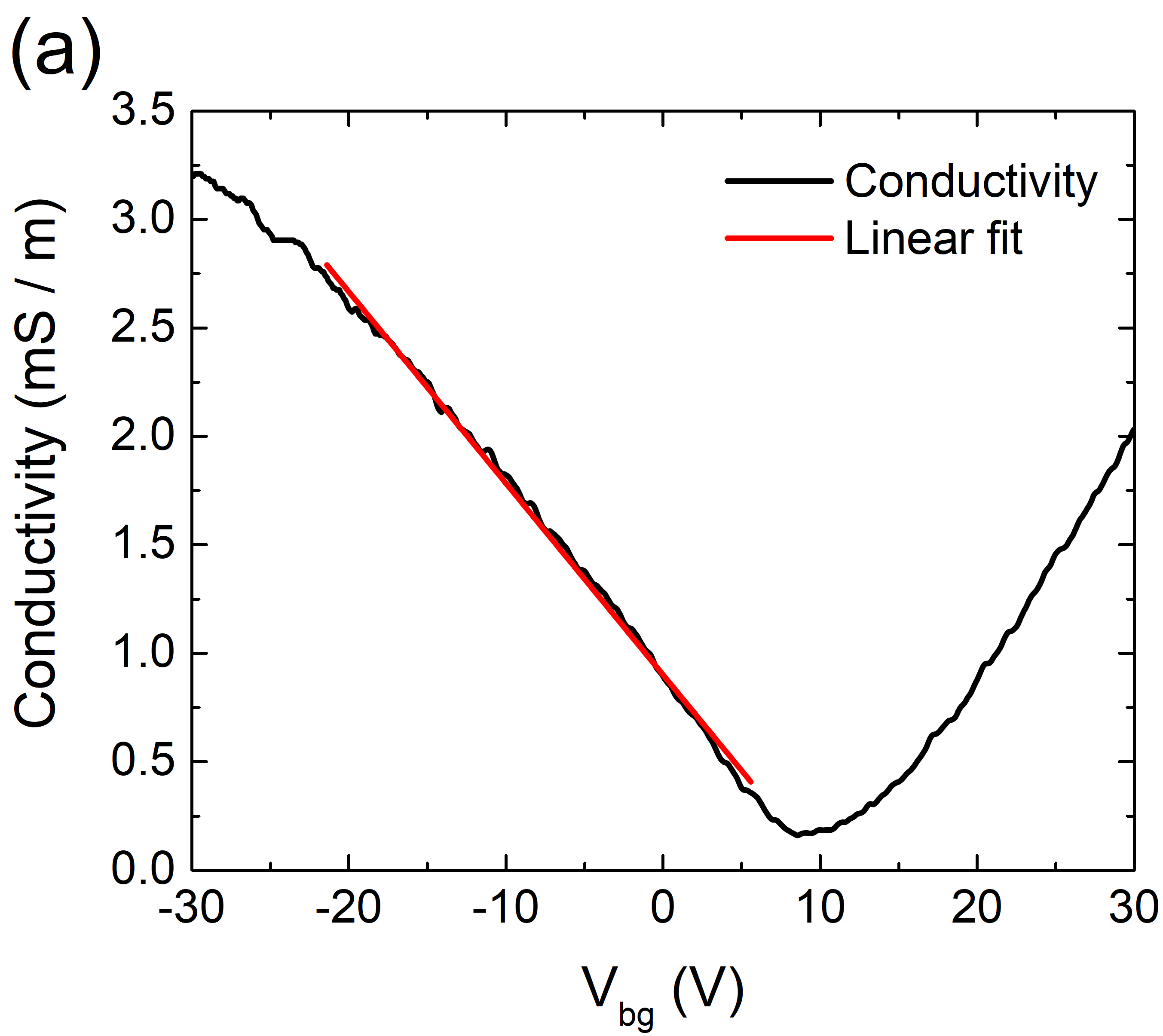}
	\includegraphics[width=0.7\columnwidth]{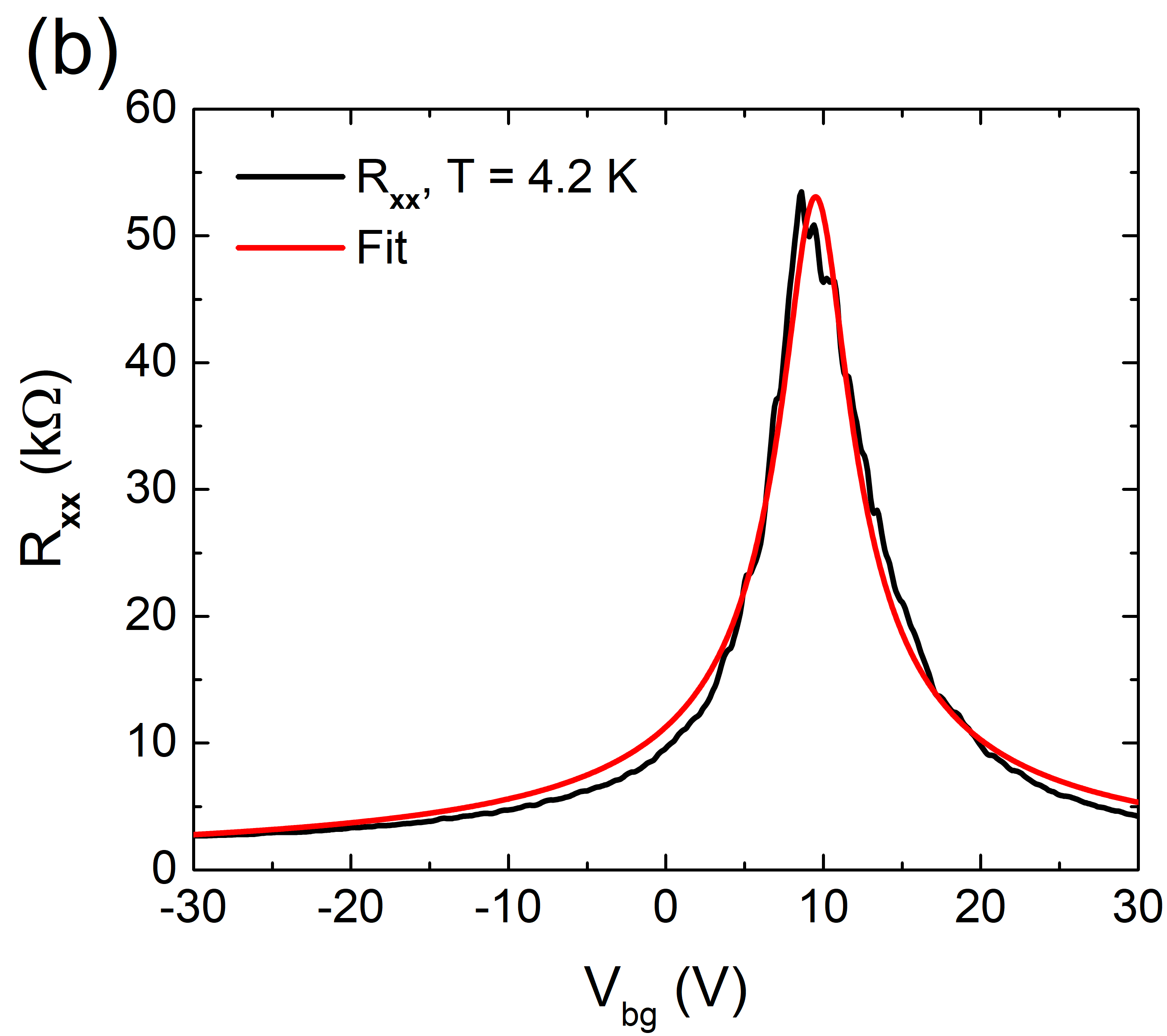}
	\caption{\label{fig:mobility}(a) The conductivity of the device at $T =$ \unit{4.2}{\kelvin} and the linear fit that is used to extract the field effect mobility. (b) The four probe resistance of the device at $T =$ \unit{4.2}{\kelvin} and the fit that is used to estimate mobility and residual charge.}
\end{figure}

The mobility of the device is estimated via a linear fit of the slope of the conductivity $\sigma$ versus back--gate voltage $V_{bg}$ (see Fig.~\ref{fig:mobility}(a)). The conductivity is given by $\sigma = \rho^{-1} = ne\mu$, where $\rho$ is the resistivity of the graphene, $n$ is the electron (hole) density, $e$ the electron (hole) charge, and $\mu$ the electron (hole) mobility. The width and length of the Hall bar is determined via scanning electron microscopy, and a mobility of $\mu = 7.2 \times 10^3$ cm$^2$V$^{-1}$s$^{-1}$ is found.

A second way to extract the field effect mobility is through a fit of the Dirac peak,\cite{Xiang2015,Xiang2016} see Fig.~\ref{fig:mobility}(b). For details, see Appendix~\ref{dirac}. We obtain $\mu = 6.8 \times 10^3$~cm$^2$V$^{-1}$s$^{-1}$, such that both values of $\mu$ agree to a reasonable degree.
 
The Dirac peak shows only a small asymmetry, suggesting that the doping influence of the metal contacts is small. The sample exhibits a hysteresis when sweeping the backgate at low temperatures, likely due to charge trapping.\cite{Wang2010} Special care is taken to guarantee consistency of the measured results.

\begin{figure}[b]
	\includegraphics[width=0.7\columnwidth]{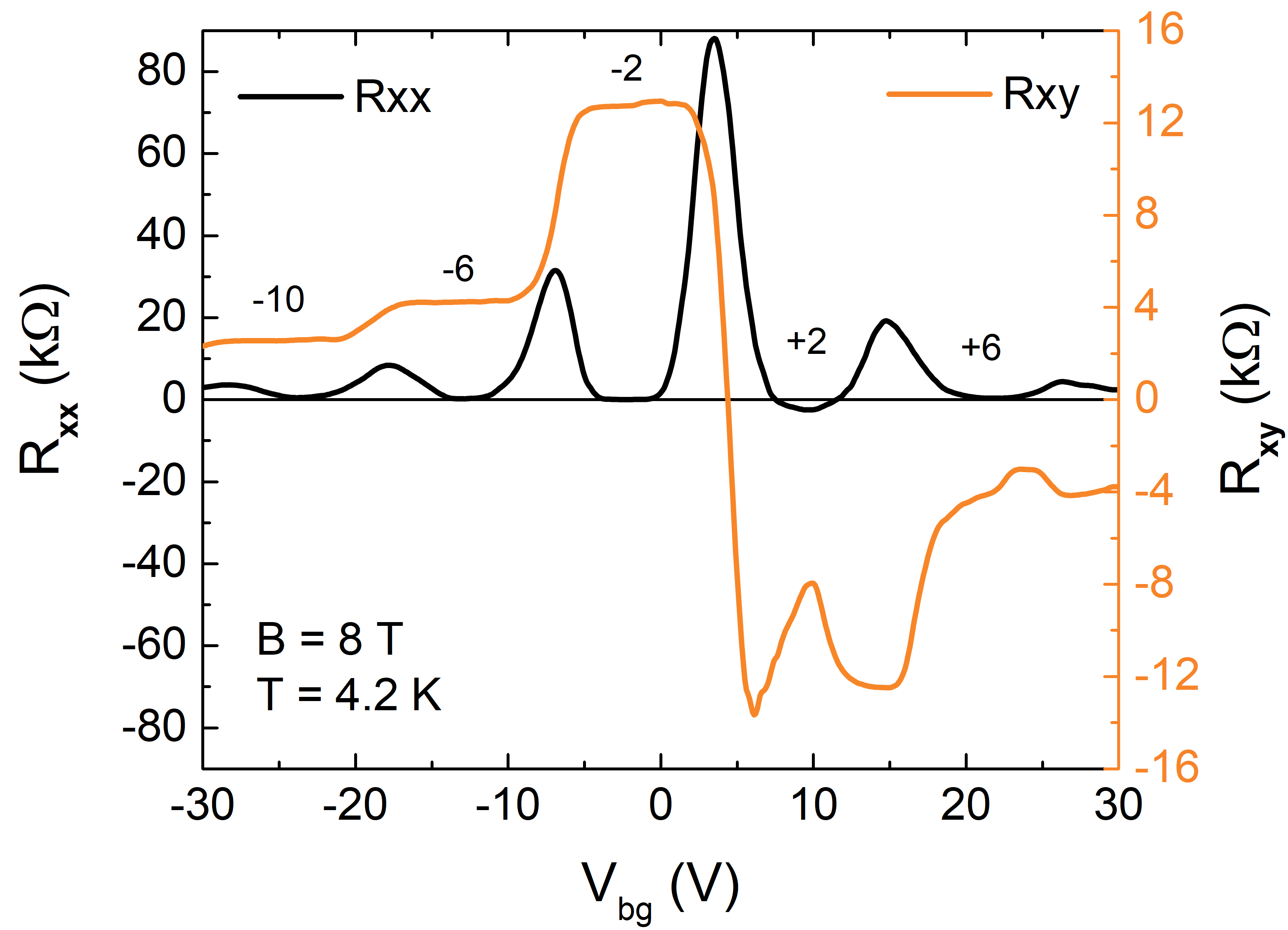}
	\caption{\label{fig:qhall}The quantum Hall effect as measured in the device, while sweeping the backgate at $T = \unit{4.2}{\kelvin}$ and $B = \unit{8}{\tesla}$.}
\end{figure}

The hole side of the sample exhibits well quantized Hall plateaus in $R_{xy}$, which coincide with minima in $R_{xx}$, as shown in Fig.~\ref{fig:qhall}. The corresponding filling factors are indicated in the figure. Here negative (positive) filling factors correspond to the hole (electron) side. Although hole and electron states in graphene are in principle equivalent, a different behavior is observed in backgate sweeps; an effect probably related to the quality of the contacts. Since the hole side show the most clear behavior we will focus on these.

\section{\label{sec:results}Experimental results}

\begin{figure}[t]
	\includegraphics[width=0.7\columnwidth]{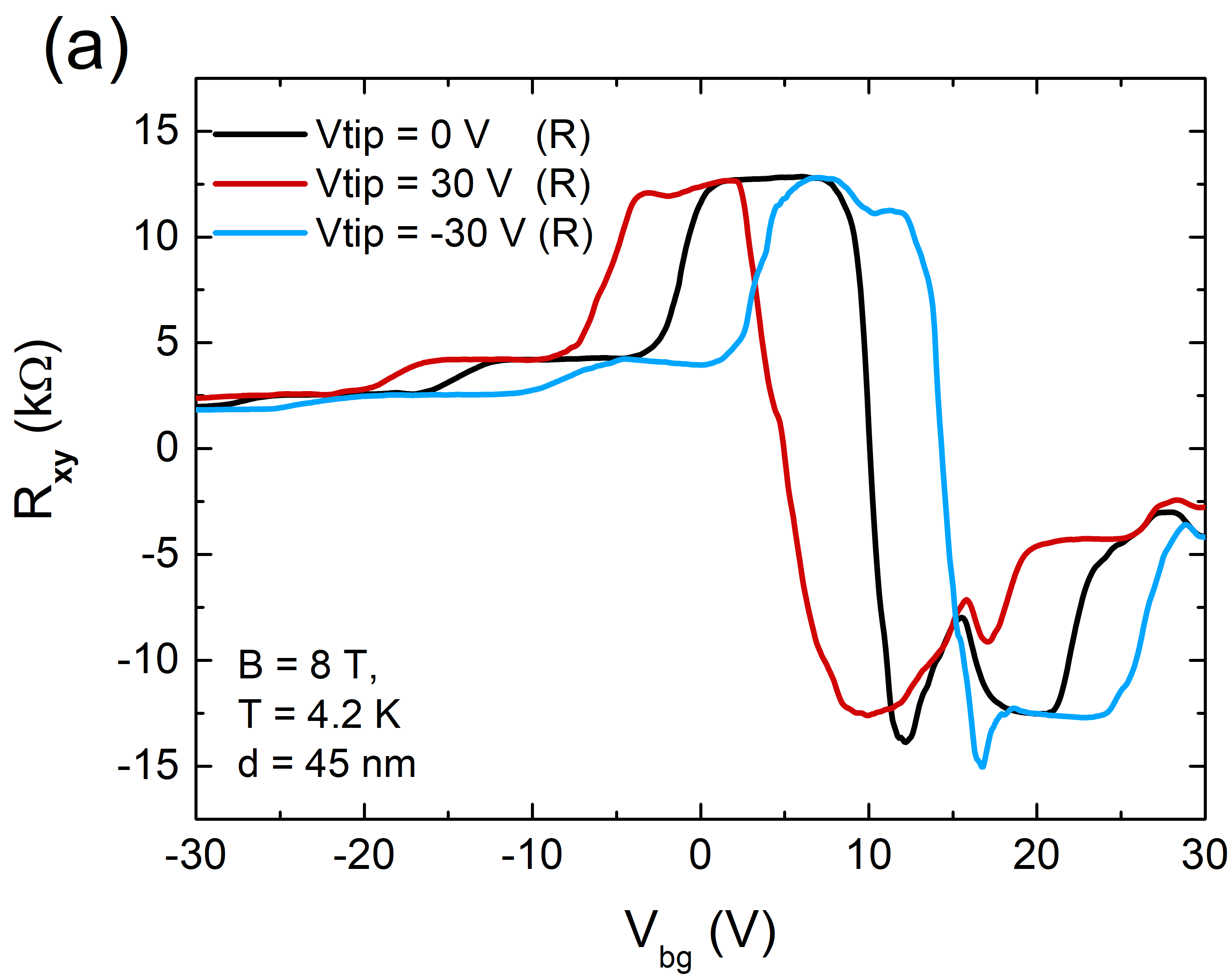}
  \includegraphics[width=0.7\columnwidth]{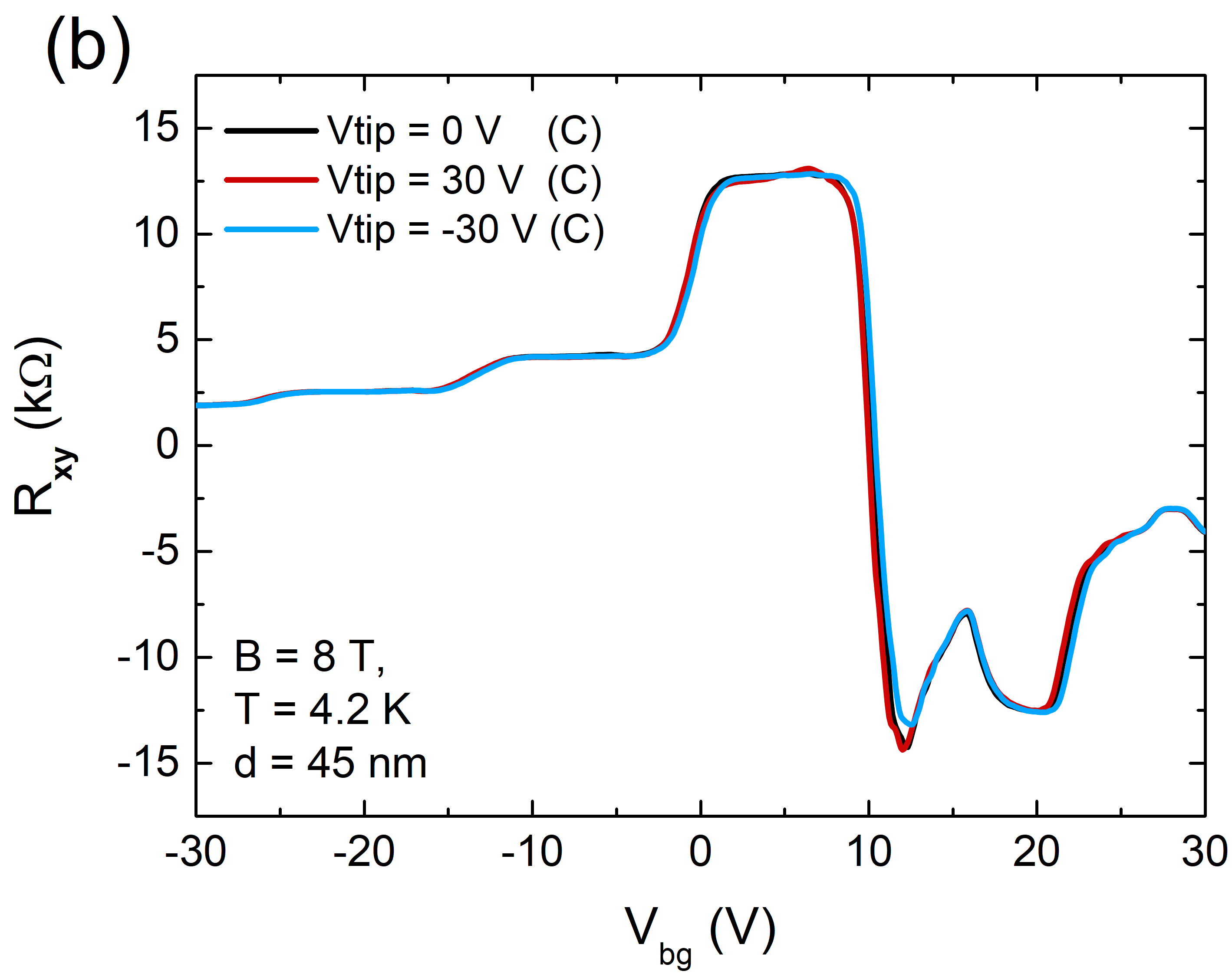}
	\caption{\label{fig:tip_effect_Rxy}(a) $R_{xy}$ vs. $V_{bg}$ with the tip placed at the right (R) side of the device, in--between the contacts used to measure $R_{xy}$. The data shows a strong dependence on the tip bias $V_{tip}$ (indicated in the legend). (b) $R_{xy}$ with the tip placed at the centre (C) of the Hall bar. $R_{xy}$ is now independent of $V_{tip}$, demonstrating that the tip effect is well localized.}
\end{figure}

To characterize the strength and the spatial extension of the tip potential, we first place the tip between two transverse contacts on the right side of the device, approximately \unit{45}{\nano\meter} above the graphene surface (as indicated by position R in Fig.~\ref{fig:sem}). The measured Hall resistance between these contacts directly reflects the filling factor in--between them ---provided that a region with a well defined filling factor extends all the way across the device--- and is not sensitive to the filling factors in other parts of the device. When a positive voltage is applied to the tip, the filling factor underneath the tip is increased, i.e.~it moves towards more positive values. An appreciable shift is observed in the $R_{xy}$ backgate sweep, when gating through the tip, as shown in Fig.~\ref{fig:tip_effect_Rxy}(a), indicating that we can indeed change the filling factor in the entire area between the contacts.

When the tip is instead placed at the centre of the device (the position marked with C in Fig.~\ref{fig:sem}), no shift in $R_{xy}$ is detected at the right side, see Fig.~\ref{fig:tip_effect_Rxy}(b). This confirms that the gating effect of the tip is sufficiently local such that it does not extend from the centre of the Hall bar to the contacts at the side, while being large enough to affect the entire width of the ribbon.

\begin{figure}[t]
	\includegraphics[width=0.85\columnwidth]{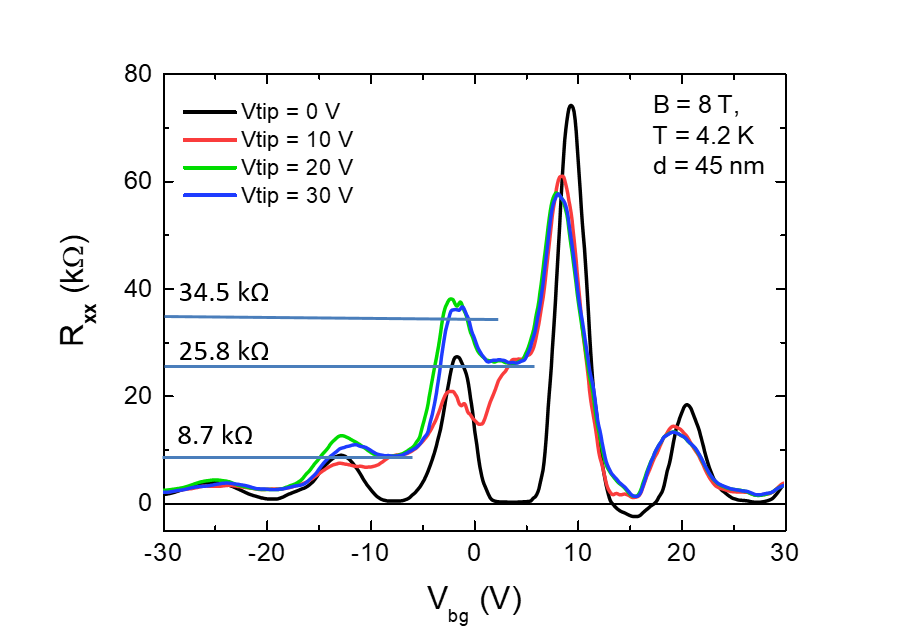}
	\caption{\label{fig:tip_centre_Rxx}The longitudinal resistance $R_{xx}$ with the tip positioned at the centre of the Hall bar, for various $V_{tip}$ biases. Several plateaus are indicated, which correspond to the backscattering of one or multiple sets of edge channels.}
\end{figure}

We now look at the value of $R_{xx}$ while the tip is at the centre of the Hall bar (Fig.~\ref{fig:tip_centre_Rxx}). When no bias is applied to the tip, $R_{xx}$ has its usual shape, cf.~Fig.~\ref{fig:qhall}. However, applying \unit{+10}{\volt} to the tip has several effects. On the hole side, two plateaus start to develop, one at $R_{xx} = \unit{8.7}{\kilo\ohm}$ and one at $\unit{25.8}{\kilo\ohm}$. These plateaus can be understood using Eqs.~(\ref{eq2})--(\ref{eq4}). Let us start with the plateau at $R_{xx} = \unit{8.7}{\kilo\ohm}$. It develops at a backgate voltage which corresponds for $V_{tip} = \unit{0}{\volt}$ to bulk filling factor $\nu = -6$. Figure~\ref{fig:tip_effect_Rxy} shows that a positive voltage on the tip moves the filling factor towards more positive values. If we assume a filling factor $\nu' = -2$ under the tip, then Eq.~(\ref{eq2}) gives $R_{xx} = 1/3 \times h/e^2 \approx \unit{8.6}{\kilo\ohm}$, in good agreement with the value of the measured plateau. Therefore, this value corresponds to a $p/p'/p$ junction with filling factors $p=-6$ and $p'=-2$. Similarly, the plateau at $R_{xx} = \unit{25.8}{\kilo\ohm}$ corresponds to a $p/n/p$ junction with filling factors $p=-2$ and $n=+2$. Then Eq.~(\ref{eq4}) gives $R_{xx} = 1 \times h/e^2 \approx \unit{25.8}{\kilo\ohm}$, again in good agreement with the measured value.

\begin{figure}[t]
	\includegraphics[width=\columnwidth]{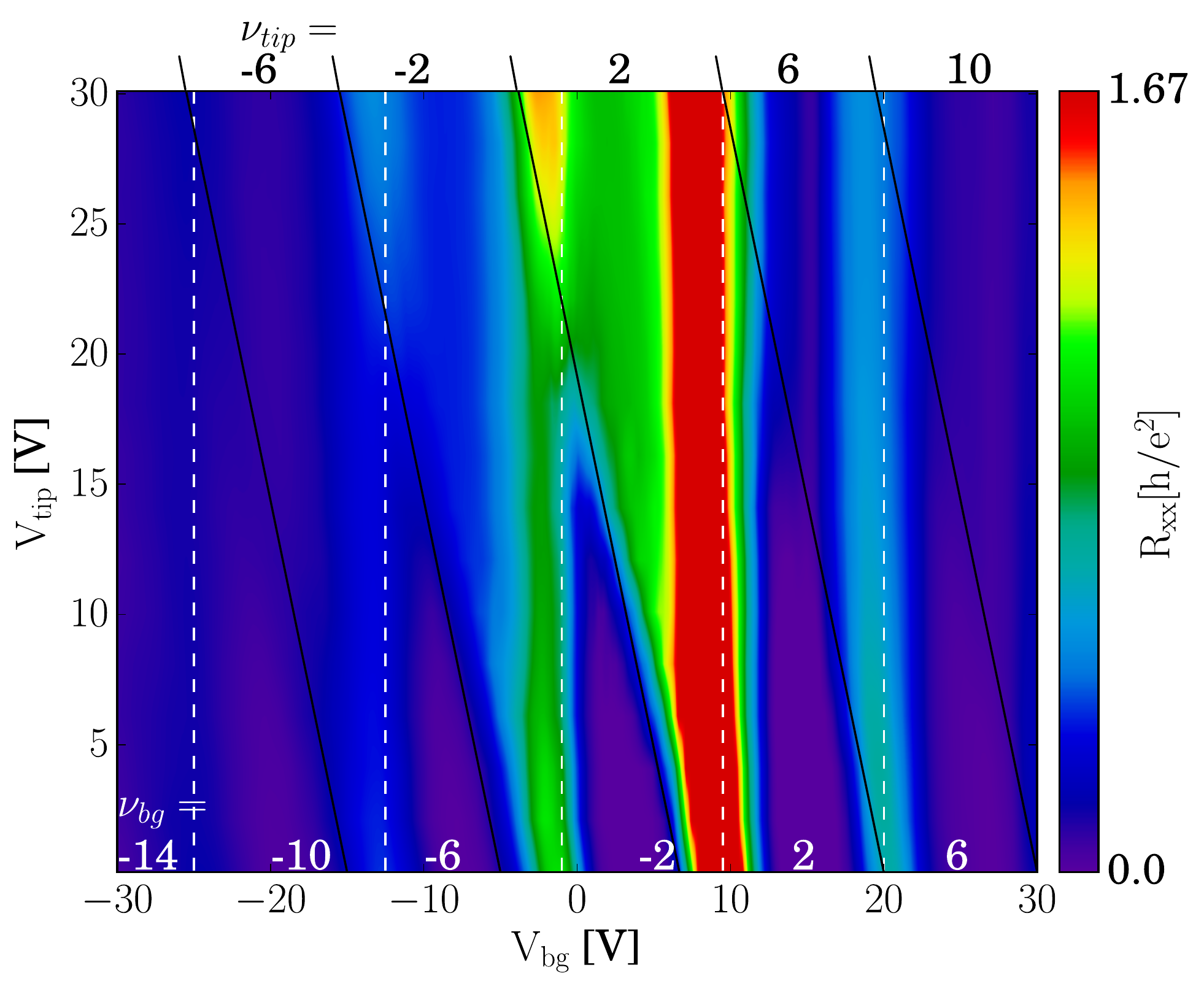}
	\caption{\label{fig:density_plot}A 2D map, showing the value of $R_{xx}$ as a function of backgate voltage $V_{bg}$ and tip voltage $V_{tip}$. The data was collected by sweeping the backgate from $-30$ to $+30$ V, while increasing the tip bias in steps of 2 V in--between sweeps, at $T = 4.3$ K and $B = 8$ T. The global filling factors $\nu_{bg}$ and the filling factors underneath the SGM tip $\nu_{tip}$ are indicated.}
\end{figure}

When the bias on the tip is increased to \unit{+20}{\volt}, the aforementioned plateaus develop further. Moreover, a new plateau with $R_{xx} = \unit{36}{\kilo\ohm}$, which is close to $4/3 \times h/e^2 \approx$ \unit{34.5}{\kilo\ohm}, appears. There is little difference between the values of $R_{xx}$ for \unit{+20}{\volt} and \unit{+30}{\volt} on the tip, indicating that the observed plateaus are robust features. The development of plateaus can be better seen in Fig.~\ref{fig:density_plot}, which shows backgate sweeps taken at many values of $V_{tip}$ in \unit{2}{\volt} intervals.

The plateaus on the right (electron) side of the curve, that would correspond to indirect backscattering in a $n/n'/n$ junction such as $n=+2$ and $n'=+6$ with $R_{xx} \approx \unit{8.7}{\kilo\ohm} = 1/3 \times h/e^2$, are notably absent. It is possible that these would--be plateaus are obscured by disruptions at the electron side as seen in Fig.~\ref{fig:qhall}.

The 2D plot reveals an interesting feature: the development of a new plateau with a longitudinal resistance value of $R_{xx} \approx 4/3 \times h/e^2$ which was also seen in the $R_{xx}$--traces shown in Fig.~\ref{fig:tip_centre_Rxx}. In order to understand this feature we used a quantum transport model which is discussed in the following Section \ref{sec:simul}.

\section{\label{sec:simul}Quantum transport modeling}

\begin{figure}[b]
        \includegraphics[width=\columnwidth]{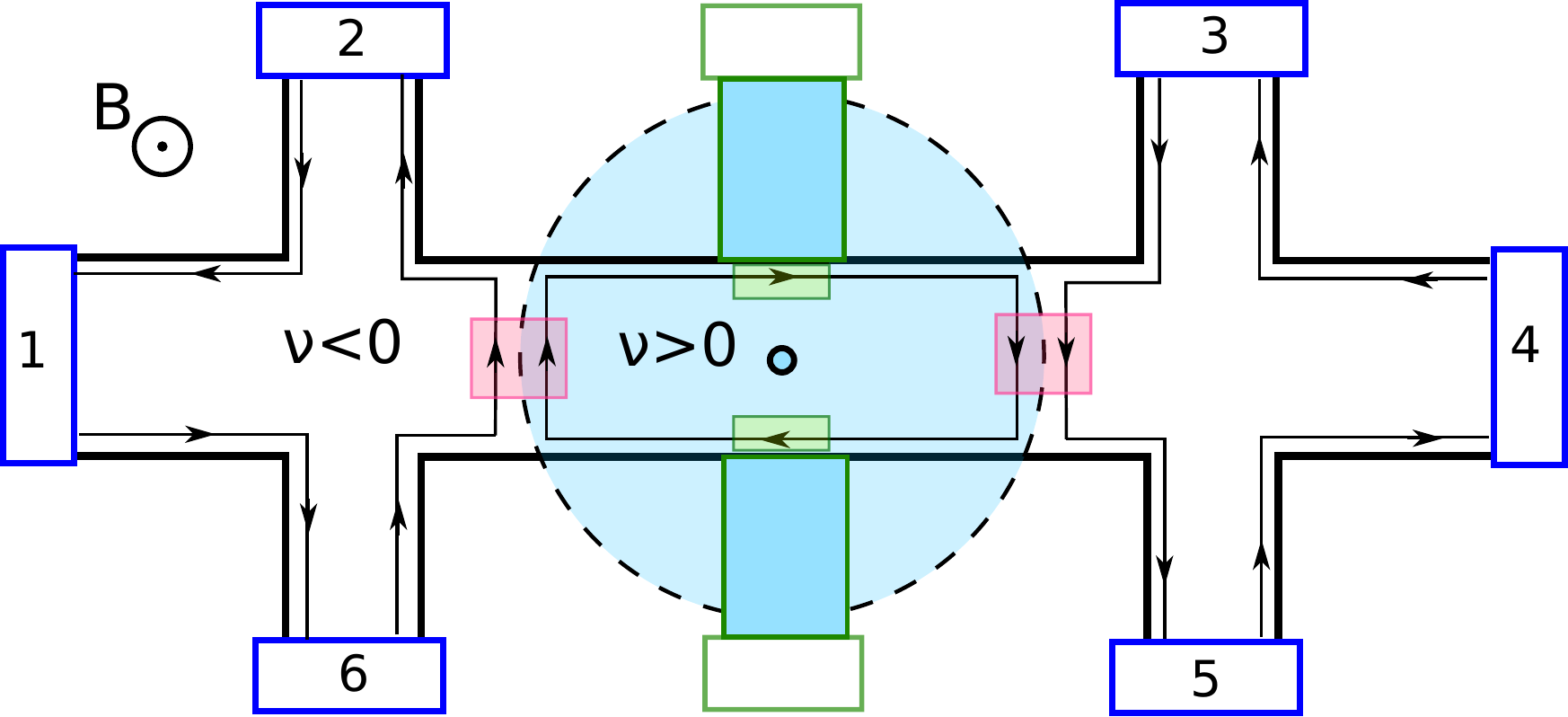}
        \caption{A schematic illustration of the system with six leads and four B\"uttiker probes. The tip is positioned at the center of the ribbon, and the strength of the tip potential is indicated by the blue circles. The tip induces a bipolar junction in the nanoribbon.
The equilibration is along the edge between channels $ \nu>0 $ as marked by green rectangles, and along the $n/p$ junction between channels  $ \nu >0 $ and   $ \nu <0 $, as marked by pink rectangles.}
        \label{fig:probes_scheme}
\end{figure}

\begin{figure*}[t]
        \includegraphics[width=0.8\textwidth]{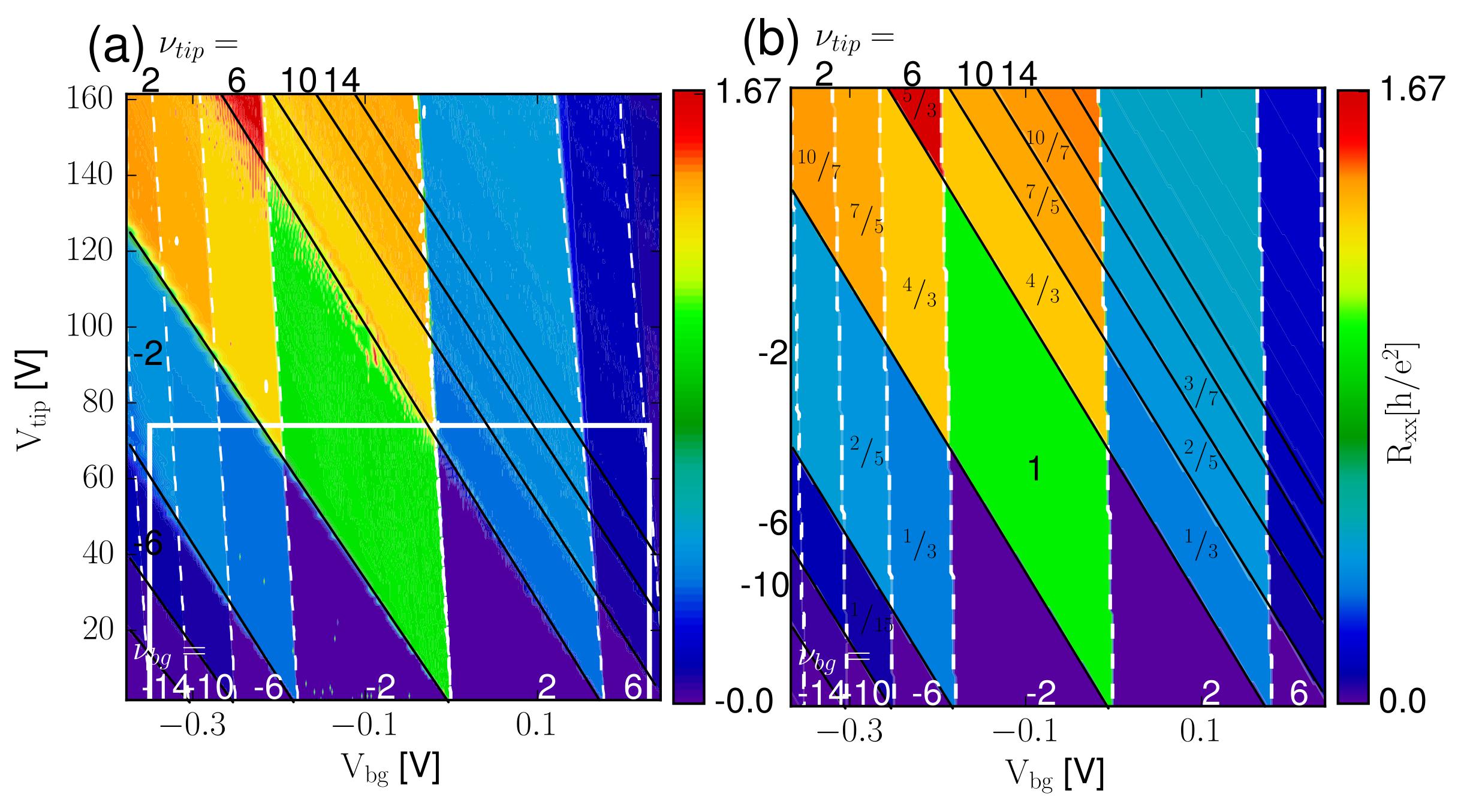}
        \caption{(a) The simulated $R_{xx}$ as a function of back gate voltage $V_{bg}$ and tip voltage $V_{tip}$. The dashed white lines show where the global filling factor $\nu_{bg}$ change, and the solid black lines where the filling factor under the tip $\nu_{tip}$ change. (b) $R_{xx}$ values calculated form Eqs. (\ref{eq:direct_equilibration}) -- (\ref{eq:both_equilibration}). The factors label the resistance values at the plateaus.}
        \label{fig:simulation_plot}
\end{figure*}

For the calculations we solve the Laplace equation for the device and the SGM tip (for details, see Appendix~\ref{details}), and for the obtained potential landscape we solve the quantum transport problem numerically. We use the tight binding formalism, with the Hamiltonian 
\begin{equation}
   H=\sum_{\langle i,j\rangle }\left( t_{ij} c_i^\dagger c_j+H.c. \right)+\sum_i V({\bf r}_i) c_i^\dagger c_i, 
\label{eq:dh}
\end{equation}
where the first sum runs over nearest neighbors, $V({\bf r}_i)$ is the external on-site potential at the position ${\bf r}_{i}$ of the $i$th atom.

Modeling of the real devices with the atomistic Hamiltonian has a large computational cost. In order to minimize it, we use the scaling approach,\cite{Rickhaus} with the scaled lattice constant $a=a_0 s_f$ and $t_{ij}=t_{ij}^0/s_f$, where the scaling parameter is taken as $s_f=4$, and the unscaled parameters are $t^0=-2.7$ eV and $a_0=2.46$ {\AA}. For the scaled system, we use  $t_{ij}=t \exp\left( \frac{2\pi i}{\phi_0}  \int_{\mathbf{r}_i}^{\mathbf{r}_j} \mathbf{A}\cdot\mathbf{dl} \right) $, with the hopping parameter $t$ and the flux quantum $\phi_0=\frac{h}{e}$.

We model the Hall-bar device as a nanoribbon, with vertical armchair and horizontal zigzag edges. The zigzag nanoribbon has a length of 980~nm and a width of 200~nm (470 atoms across the ribbon) and is connected to narrow armchair--type leads used as voltage probes, of width 69~nm (140 atoms across the ribbon) and length 122~nm, which are labeled $2-3$ and $5-6$ in Fig.~\ref{fig:probes_scheme}. The model system is obtained by scaling down the experimental device by another factor of $\sim$ 4. We therefore use $B=32$~T.

\begin{figure}[t]
        \includegraphics[width=\columnwidth]{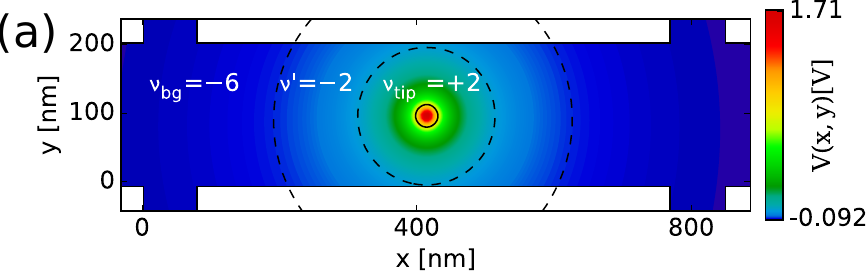}
				\includegraphics[width=\columnwidth]{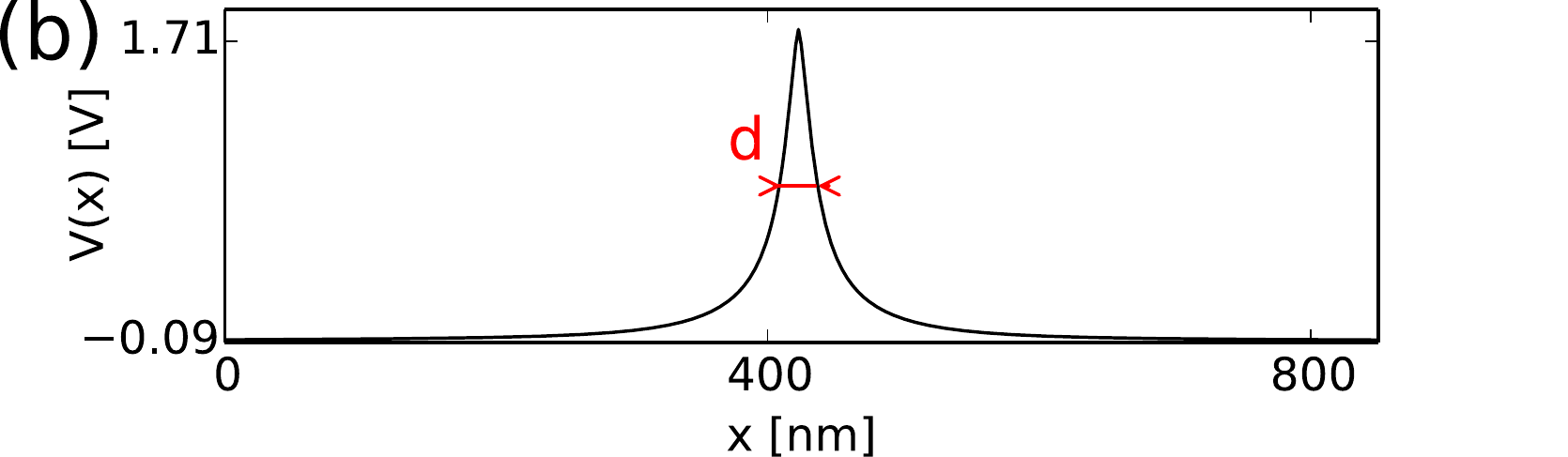}
        \includegraphics[width=\columnwidth]{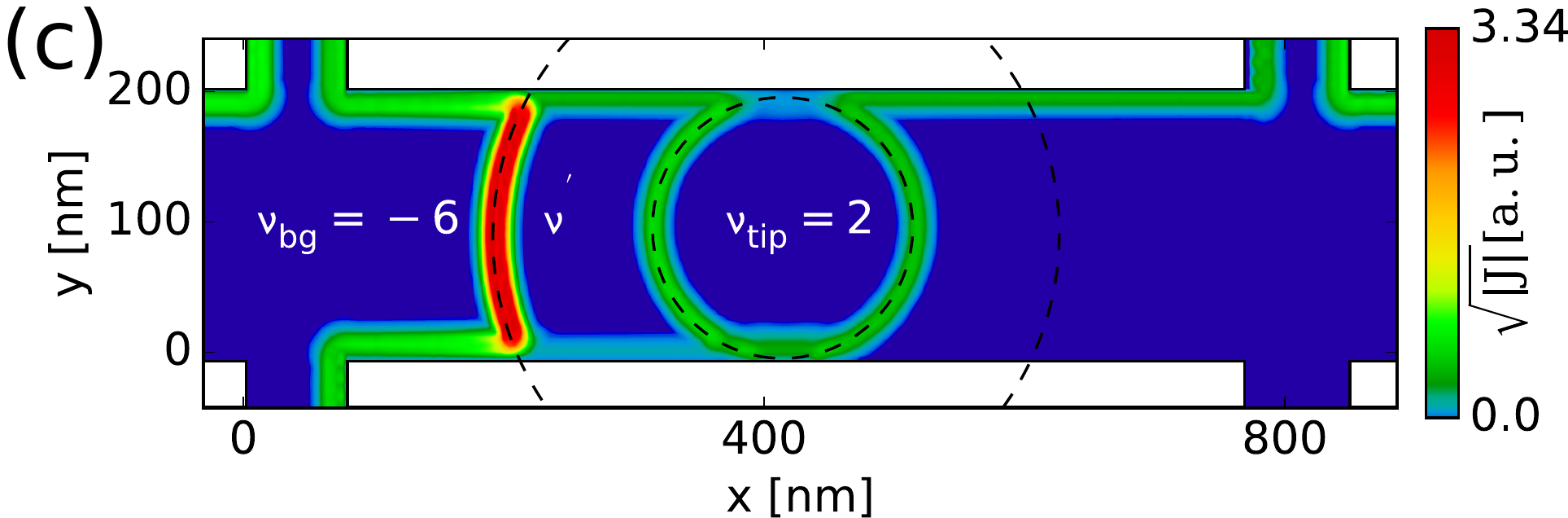}
        \caption{(a) A representative potential map of the system for $ \nu_{bg} = -6 $, $ \nu_{tip} = +2 $. The dashed circles show where the potential equals the Fermi energy (inner circle) and the energy of the 1st LL (outer circle). The solid circle indicates the potential at half maximum. (b) Horizontal cross--section through the center of the Hall--bar in (a), showing how the potential varies with position. (c) An exemplary current density in the system for the potential profile in (a).}
        \label{fig:current_43}
\end{figure}

A representative potential profile obtained from the numerical solution of the Laplace equation is presented in Fig. \ref{fig:current_43}(a,b). The width $d$ at half-maximum of this potential is 30~nm, and the diameter of the induced n-p junction is 200.4~nm which is comparable to the nanoribbon width 200~nm, whereas the magnetic length for the modeled system is $l_B=\sqrt{\tfrac{\hbar}{eB}}=4.5$~nm. The Laplace equation scales linearly with the system size, and we can estimate that in the real sample, which is about 4 times bigger, for the gate voltages that result in same filling factors, the width at half-maximum is 120~nm, whereas for $B=8$~T used in experiment the magnetic length is 9~nm. In comparison, the typical extent of the smoothed potential induced by a standard top-gate in graphene lying on SiO$_2$ substrate is of the order of several tens of nanometers \cite{Macucci2015}.

For the solution of the transport problem, we use the Landauer-B\"uttiker approach \cite{Buttiker1986} and calculate the resistance of the sample from the solution of the quantum scattering problem. For details, see Appendix~\ref{details}. We assume zero temperature. In the numerical model, equilibration is obtained with the dephasing introduced by virtual B\"uttiker probes.\cite{ButtikerProbes} We place the B\"uttiker probes in two locations: (1) At the $n/p$ junctions present in the system. They are located in the two spots marked by pink rectangles in Fig.~\ref{fig:probes_scheme}. (2) At the sides of the central part of the nanoribbon, one at the upper edge and one at the bottom edge of the ribbon, indicated in green in Fig.~\ref{fig:probes_scheme}. For details on the B\"uttiker probes, see Appendix~\ref{details}. The probes (1) induce equilibration of the co--propagating channels  along the bipolar junction, while the probes (2) mix the Landau levels at the edge of the ribbon.

The colormap in Fig.~\ref{fig:simulation_plot}(a) presents the calculated results for the longitudinal resistance of the device as a function of back gate voltage $V_{bg}$ and tip voltage $V_{tip}$ obtained with the method described above. The white box indicates the area to be compared with the experimental results of Fig.~\ref{fig:density_plot}. A very good agreement is obtained for the resistance values at the plateaus. The filling factors under the tip, $\nu_{tip}$, and far from it, $\nu_{bg}$, obtained from the simulations, are given in Fig.~\ref{fig:simulation_plot}(a). For example, for the case of $\nu_{bg} = -6$ and $\nu_{tip} = -2$, we obtain $R_{xx} = 1/3 \times h/e^2 \approx \unit{8.6}{\kilo\ohm}$, in good agreement with the interpretation of the value of the measured plateau in Fig.~\ref{fig:tip_centre_Rxx}. Similarly, for $\nu_{bg} = -2$ and $\nu_{tip} = +2$, we obtain $R_{xx} = 1 \times h/e^2 \approx \unit{25.8}{\kilo\ohm}$. This good agreement allows to index the filling factors also for the experimental data shown in Fig. \ref{fig:density_plot}. There and in Fig.~\ref{fig:simulation_plot}(a) we find that the plateau with the resistance value close to $4/3 \times\frac{h}{e^2} \approx 34.5$~k$\Omega$ that in the experiment has started to develop corresponds to a configuration of filling factors $\nu_{bg}=-6$ and $\nu_{tip}=+2$. 

The calculated current density plot obtained by solving the scattering problem for this particular condition is presented in Fig.~\ref{fig:current_43}. One can see two distinct current branches belonging to hole Landau levels number 0 (the circular branch) and 1 (the leftmost branch) that are separated by an area of an intermediate filling factor $\nu'=-2$. The intermediate region results from the fact that the potential profile of the point--like tip is smooth. Due to the smoothness of the potential, the LLs with $|N|>0$ are spatially separated from the 0th LL, thus along the $n/p$ junction the equilibration takes place only between the lowest Landau level channels,\cite{Klimov2015,gasse2016} in contrast to sharp bipolar junctions where the equilibration is between all modes.
 
Beyond the range covered in the experiment, we find also a plateau of similar $R_{xx}$ for $\nu_{bg}=-2$ and $\nu_{tip}=+6$. Other new plateaus occur, e.g. $R_{xx}\approx1.4\times\frac{h}{e^2}$  at  $\nu_{bg}=-10$ and $\nu_{tip}=+2$. Generally, the resistance values are symmetric with respect to the simultaneous exchange of $\nu_{bg}\rightarrow-\nu_{tip}$ and $\nu_{tip}\rightarrow-\nu_{bg}$.

\section{\label{sec:discuss}Discussion}

\begin{figure}[t]
        \includegraphics[width=\columnwidth]{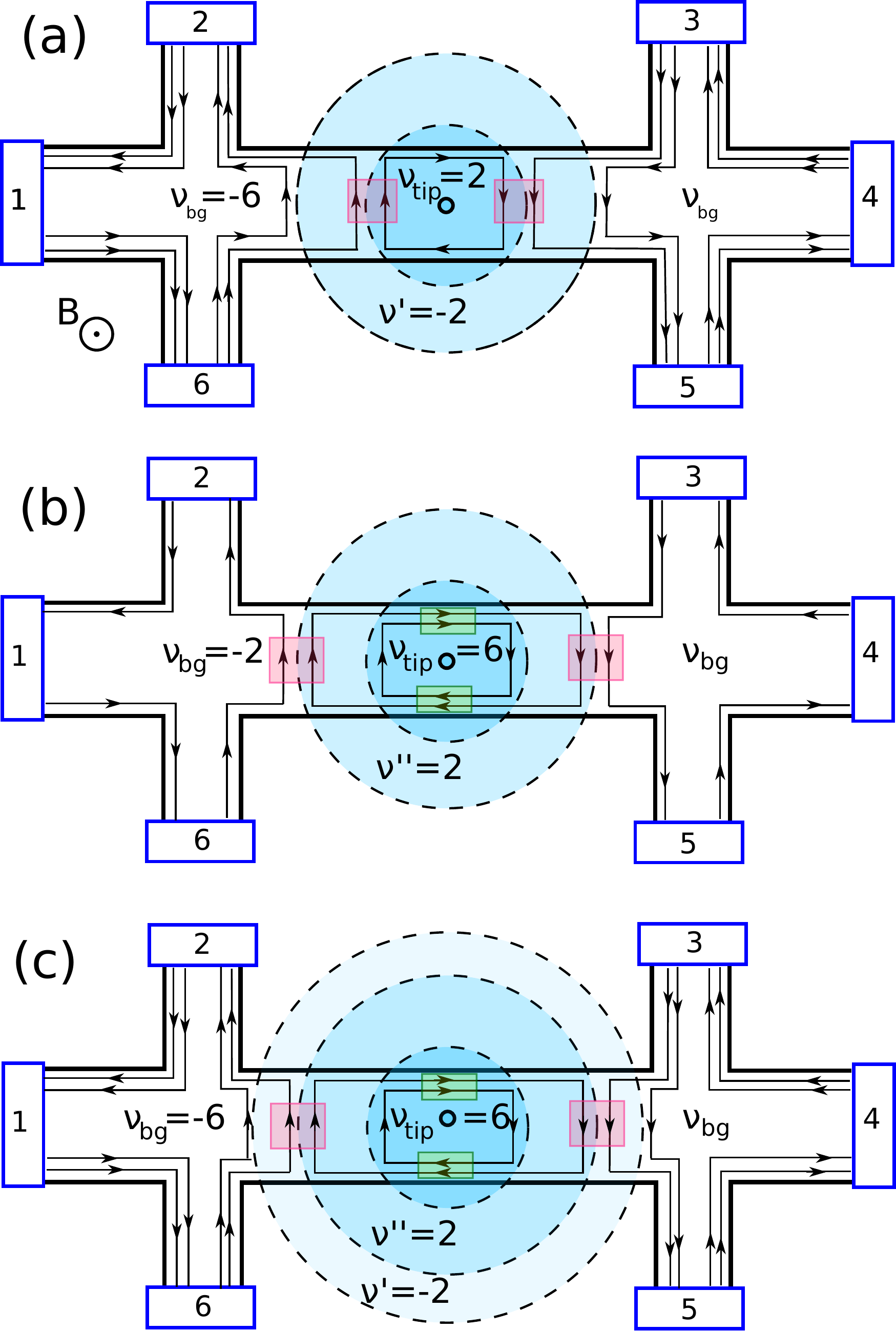}
        \caption{(a) A schematic illustration of the backscattering scenario of the lowest Landau level channels only. The tip is positioned at the center of the ribbon, and the strength of the tip potential is indicated by the blue circles. In this scenario the outermost channels $\nu_{bg}$ (here $\nu_{bg} = -6$) are backscattered and do not equilibrate with the innermost channels $\left( \nu_{tip} = +2 \right)$. (b) Backscattering of two sets of edge channels. In this scenario the innermost channels $\left ( \nu = +6 \right)$ do not equilibrate with the outermost channels $\left( \nu = -2 \right)$. The equilibration takes place along the edge between channels $ \nu = +6 $ and   $ \nu = +2 $, marked by green rectangles, and along the $n/p$ junction between channels  $ \nu =-2 $ and   $ \nu = +2 $, as marked by pink rectangles. (c) A schematic illustration of the general backscattering scenario. Here, the channels $\nu'$ and $\nu''$ equilibrate along the $n/p$ junction, whereas the outermost channels $\nu_{bg}$ are backscattered and the innermost ones $\nu_{tip}$ undergo an indirect equilibration along the edge.}
        \label{fig:backscattering}
\end{figure}

The newly observed plateaus deserve a discussion. We use the following notation: $\nu'$ is used for a filling factor outside the bipolar junction, which is different from $\nu_{bg}$. $\nu"$ is used to refer to a filling factor inside the bipolar junction, which differs from $v_{tip}$.

Under the assumption that equilibration takes place only between the lowest Landau level channels, we can derive expressions for the resistance in case of ideal equilibration, using the Landauer-B\"uttiker approach. By analysis of the filling factors, we find two regimes of backscattering. In the first case, presented schematically in Fig.~\ref{fig:backscattering}(a), only equilibration along the $n/p$ junction is present. In this case, the longitudinal resistance, as calculated analytically in Appendix~\ref{matrix}, is
\begin{equation}
R_{xx} = \frac{h}{e^2} 
\left( \frac{ 2}{|\nu'|} +  \frac{ 1}{|\nu_{tip} |} -   \frac{ 1}{|\nu_{bg} |} \right) .
\label{eq:direct_equilibration}
\end{equation}
Assuming that only $|\nu'|=2$ and $|\nu_{tip}|=2$ equilibrate, we obtain for $|\nu_{bg}|=6,10,14,...$ a series of fractions $R_{xx}=\frac{4}{3}, \frac{7}{5}, \frac{10}{7},... \times \frac{h}{e^2}$. In particular, for $\nu_{bg}=-6$ and $\nu_{tip}=+2$, the fraction  $R_{xx}=\frac{4}{3} \times \frac{h}{e^2}$ is clearly close to the value obtained in the experiment and in the transport calculations. Also, for $\nu_{bg}=-10$ and $\nu_{tip}=+2$, there is $R_{xx}=\frac{7}{5} \times \frac{h}{e^2}$, in perfect agreement with the modeled  $R_{xx}=1.4 \times \frac{h}{e^2}$.

The second case is schematically presented in Fig.~\ref{fig:backscattering}(b), with equilibration both along the $n/p$ junction and along the edges between channels $\nu''$ and $\nu_{tip}$. In this case, the resistance reads
\begin{equation}
R_{xx} = \frac{h}{e^2} 
\left( \frac{ 2}{|\nu''|} - \frac{ 1}{|\nu_{tip} |} +  \frac{ 1}{|\nu_{bg} |} \right) ,
\label{eq:indirect_equilibration}
\end{equation}
which, for the case $|\nu_{bg}|=2$ and $|\nu''|=2$, for $|\nu_{tip}|=6,10,14,...$ gives $R_{xx}=\frac{4}{3}, \frac{7}{5}, \frac{10}{7},... \times \frac{h}{e^2}$.

The generalization of the two cases is a scenario presented schematically in Fig.~\ref{fig:backscattering}(c). 
In this case, the resistance is 
\begin{equation}
R_{xx} = \frac{h}{e^2} \left( \frac{2}{ |\nu'| } + \frac{2}{ |\nu''| } - \frac{1}{ |\nu_{tip}| }  - \frac{1}{ |\nu_{bg}| } \right) .
\label{eq:both_equilibration}
\end{equation}
Here the equilibration is between the channels $\nu'$ and $\nu''$ along the $n/p$ junction, the outermost channels $\nu_{bg}$ are backscattered, and the innermost ones $\nu_{tip}$ equilibrate with the channel $\nu''$ along the edge. In particular, for the example given in Fig.~\ref{fig:backscattering}(c), $\nu_{bg}=-6$ and $\nu_{tip}=+6$, $\nu'=-2$ and $\nu''=+2$, Eq. (\ref{eq:both_equilibration}) gives $R_{xx}=\frac{5}{3} \times \frac{h}{e^2}$, which is close to the simulated value (see Figs.~\ref{fig:simulation_plot}(a) and (b)).

The parallelogram of the analytical values of longitudinal resistances is shown in Fig.~\ref{fig:simulation_plot}(b), with several plateaus labeled by fractions which equal the analytically derived $R_{xx}$. It shows good agreement with the modeled values in Fig.~\ref{fig:simulation_plot}(a). This confirms our hypothesis that the higher channels are well separated and backscattered without being equilibrated.

For graphene encapsulated in hBN, due to the suppression of disorder, in high magnetic field spin-splitting can be observed, which gives rise to suppression of inter-channel scattering at filling factor $|\nu|$=1 \cite{Weie2017,Amet2014}. In our study, the Zeeman splitting is 0.92 meV, no spin-splitting is observed, and only plateaus at even filling factors are resolved in Fig.~\ref{fig:qhall}. The would-be odd filling factor plateaus are obscured by the smooth transition between the even plateaus due to the substrate-induced disorder. Accordingly, in the numerical calculations we assume spin degeneracy for all Landau levels.

\section{\label{sec:summary}Summary}

We have demonstrated backscattering of quantum Hall edge channels in graphene through local gating achieved by the tip of a Scanning Gate Microscope. Moreover, due to the potential generated by the SGM, we observe a gradual change of the filling factor along the device, which culminates in a  fractional value of $R_{xx}$ that has not been reported before. The underlying processes are well understood, and these results are supported by tight--binding simulations. 

Being able to move the tip freely gives great freedom in manipulating the edge channels, which can be done by changing the tip position, tip--sample distance, and the applied voltage. Secondly, not having to fabricate additional split--, buried--, or top--gates simplifies the fabrication process and avoids possible detrimental effects on the quality of the graphene.

The work presented here paves the way for future experiments in which one takes full advantage of the flexibility of the SGM, by illustrating that this type of experiments is within experimental reach. An interesting extension to this work would be, among others, to study the transition from well separated via interacting to (fully) backscattered edge--channels, by gradually approaching the device from the side with a biased tip.

We believe that such an experiment can shed light on the elusive microscopic structure of the quantum Hall edge channels and the process of electrostatic reconstruction.\cite{Chklovskii1992,Chklovskii1993,Chamon1994} Furthermore, one can further study such dynamics by combining the SGM with a split--gate. Such a device would also offer the possibility of conducting interference experiments.

\appendix

\section{\label{dirac}Fit of the Dirac peak}

The induced charge is estimated by treating the graphene as one plate of a parallel plate capacitor, and the backgate as the other. In this case the dielectric filling between the two plates is the \unit{300}{\nano\meter} thick layer of SiO$_2$, with $\varepsilon \approx 3.9$. Given the capacitance $C$, the induced charge is 
\begin{eqnarray}
n & = & \frac{C}{e} \left| V_{bg} - V_{cnp} \right| = \frac{11.5}{1.6 \times 10^{-19}} \frac{\left[ \mu F / cm^2 \right]}{\left[ C \right]} \nonumber \label{eq5} \\
  & = & 7.2 \times 10^{10} \left[ V^{-1} cm^{-2} \right] \times \left| V_{bg} - V_{cnp} \right|,
\end{eqnarray}
which leads to the following expression for the resistance R which is used to fit the data:
\begin{equation}
R = \frac{L}{W} \left( e \mu \sqrt{n_0^2 + \left( 7.2 \times 10^{10} \times \left( V_{bg} - V_{cnp} \right) \right)^2}\right)^{-1}. \label{eq:fit}
\end{equation}
$n_0$ is called the `left over' or residual charge and represents the charge inhomogeneity at the charge neutrality point (CNP). 
From the fit, the parameters $\mu = 6.8 \times 10^3$~cm$^2$V$^{-1}$s$^{-1}$ and $n_0 = 1.5 \times 10^{11}$~cm$^{-2}$ are extracted. At zero backgate voltage, the carriers are p--type, and have a density of $n \approx 8.7 \times 10^{11}$~cm$^{-2}$.

\section{Details of the Quantum Transport Model\label{details}}

The electrostatic potential $V(x)$ is obtained by solving the Laplace equation on a three dimensional mesh of dimensions in $x\times y\times z$ given by $ 1972\times 820\times 2404 $~nm$^3$. We assume the SiO$_2$ spanning between $z=0$ and $z=84$~nm, graphene lying on top of the dielectric, and the tip at $z=96$~nm. The tip is modeled as a point charge at the center of the Hall--bar, 12~nm above the graphene. We assume boundary conditions given by $V=V_{bg}$ at the bottom of the computational box, $V=V_{tip}$ at one point of the mesh (12~nm above the graphene, in the central $x-y$ position), and the condition for zero normal component of electric field at all the computational box walls except the bottom one. At the interface between SiO$_2$ (dielectric constant $\varepsilon_1 = 3.9$) and vacuum ($\varepsilon_1 = 1$) we use the boundary condition $\varepsilon_1 E_1 = \varepsilon_2 E_2$, where $E_1$ ($E_2$) is the electric field below (above) the graphene.

Once the Laplace equation is solved for the system, we obtain a potential landscape which allows us to determine the filling factor in the area under the tip and further away from the tip, given the energies of Landau levels in the nanoribbon. The latter is determined from the solution for the modes in a nanoribbon in the external magnetic field.

In the fully coherent transport problem, equilibration \cite{Abanin2007,Ozyilmaz2007,Ki2009} of the channels does not occur, and no resistance plateaus similar to the experimental ones can be obtained. In the numerical model the equilibration is introduced with the dephasing by virtual B\"uttiker probes  \cite{ButtikerProbes}. In these probes a zero net current is set, i.e.~electrons which enter the probe, equilibrate in the reservoir and re--enter the system with a random phase. We obtain an agreement with the experimental $R_{xx}$ diagram setting the B\"uttiker probes in two locations: (1) At the $n/p$ junctions whenever present within the system. We use the  probes connected vertically to the interior of the graphene sheet, similar as in Ref.~\onlinecite{Xiang2016}. The probes consist of 45 zigzag chains, aligned in a rectangle of width $W_y=48$~nm and length $W_x=20$~nm. They are located in two spots marked by pink rectangles in Fig.~\ref{fig:probes_scheme}, with the $y$-position in the middle of the nanoribbon, and the $x$-position  chosen at the bipolar junction, at the point where the potential is equal to the Fermi energy. (2) At the sides of the central part of the nanoribbon we put halfway along the ribbon two narrow leads, one at the upper edge and one at the bottom edge of the ribbon, which are semi--infinite in $y$ direction, armchair ribbons of the width of 68 atoms.

\subsection{Landauer-B\"uttiker formula}

We use the Landauer-B\"uttiker formalism for both quantum transport calculation and derivation of the analytical formulas for longitudinal resistance. For the scattering problem, we use the wave function matching (WFM). For the details of the method refer to Ref.~\onlinecite{Kolacha}. The transmission probability from terminal $l$ to mode $m$ in the terminal $k$ is
\begin{equation}
T_{kl}^m = \sum_{ n } \left|t_{mn}^{kl}\right|^2,
\label{eq:transprob}
\end{equation}
where $t_{mn}^{kl}$ is the probability amplitude for the transmission from the mode $n$ in terminal $l$ to mode $m$ in terminal $k$. Then the conductance from lead $l$ to $k$ is given by
\begin{equation}
G_{kl} = G_0 \sum_{ m } T_{kl}^m,
\end{equation}
where $G_0=\frac{2e^2}{h}$ is the conductance quantum.

In the general case of $N$ terminals, 
one constructs the $\boldsymbol{G}$ matrix that satisfies
$ \boldsymbol{I}=\boldsymbol{G}\boldsymbol{V} $. The diagonal elements are
\begin{equation}
\boldsymbol{G}_{ii} = \sum\limits_{j=1,j\ne i}^{N} 
G_{ij}
\label{eq:Gdiag}
\end{equation} 
and the off--diagonal ones 
\begin{equation}
\boldsymbol{G}_{ij} = - G_{ij}.
\label{eq:Goffdiag}
\end{equation} 
Due to Kirchhoff's law, the currents in all $N$ leads are not independent, thus one eliminates $I_{N}$ and assumes $V_{N}=0$. The final matrix is of dimension $(N-1)\times (N-1)$. One calculates the resistance matrix as $\boldsymbol{R}=\boldsymbol{G}^{-1}$. Assuming that the current flows between two chosen terminals $i$ and $j$, and zero net current in all other terminals, from the voltage measured between the terminals $k$ and $l$ one can obtain the resistance as $R_{ij,kl} = \left.\frac{R_k-R_l}{I_i}\right|_{I_j=-I_i}$.

Fig.~\ref{fig:probes_scheme} shows the labeling of the terminals. The longitudinal resistance $R_{xx} = R_{14,65}$, with current flowing between leads 1 and 4, and voltage drop measured between leads 6 and 5, is calculated by constructing the conductance matrix $\boldsymbol{G}$.\cite{datta} For the numeration of terminals shown in Fig.~\ref{fig:probes_scheme} we can calculate the longitudinal resistance as $R_{14,65} = R_{54}-R_{51}$.

\subsection{Derivation of longitudinal resistance values\label{matrix}}

\begin{figure}[b]
        \includegraphics[width=\columnwidth]{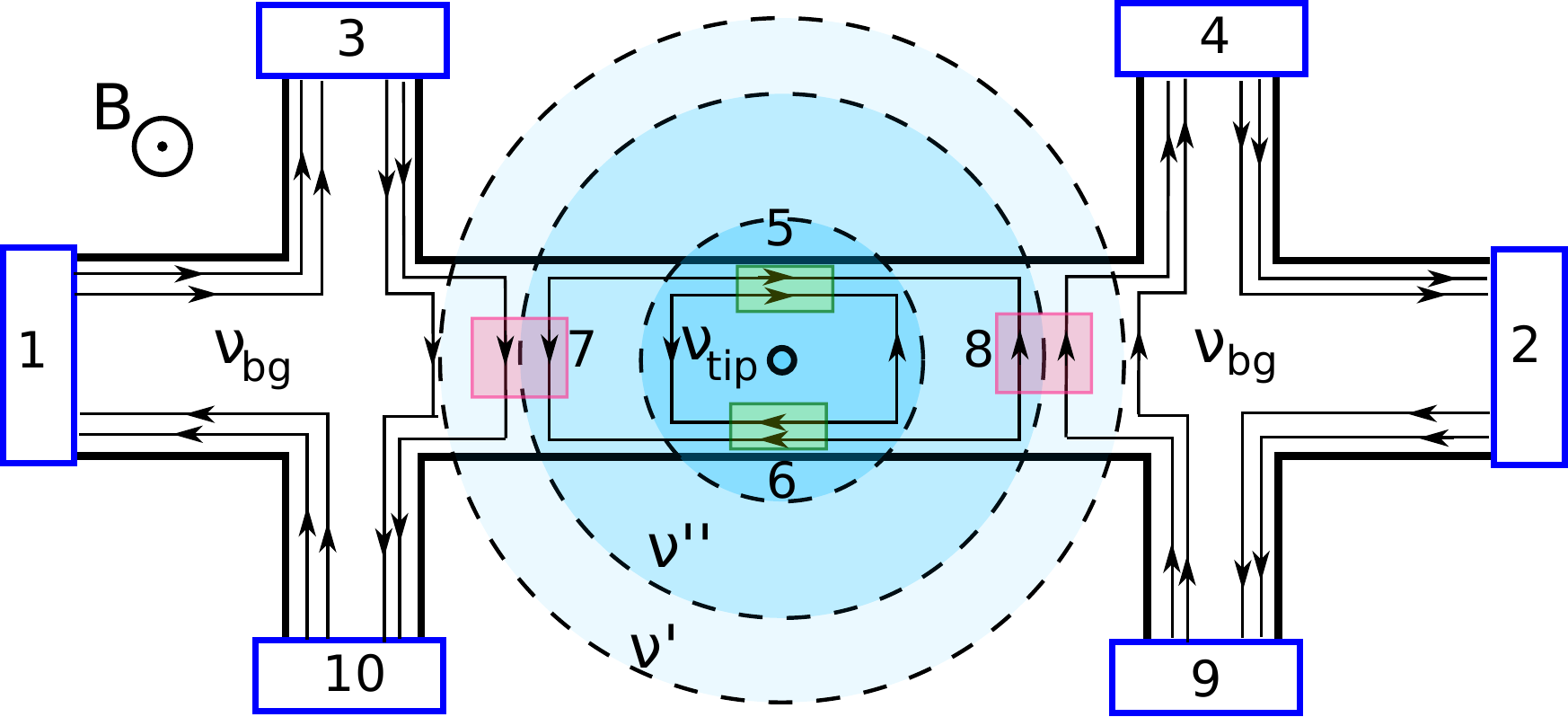}
        \caption{A schematic drawing of the system for the calculation of longitudinal resistances, with the numeration of the leads. The pink and green rectangles mark the B\"uttiker probes which absorb all the current flowing through the rectangles.}
        \label{fig:backscattering_scheme}
\end{figure}

To account for the channels equilibration, we use additional B\"uttiker probes. In the system shaped into a Hall--bar there are 6 contacts, and the B\"uttiker probes are located along the the current paths where the channels mix, as marked by pink and green rectangles in Fig.~\ref{fig:backscattering_scheme}. 4 probes are sufficient to accurately model the equilibration in the system. 
In total there are 10 terminals.
In Fig.~\ref{fig:backscattering_scheme} the numeration of the terminals used for the calculation of the analytical formulas for longitudinal resistance is shown. The matrix calculated for the case shown in Fig.~\ref{fig:backscattering_scheme} is
\begin{widetext}
\begin{equation}
\boldsymbol{G} = \frac{e^2}{h} \left(
\begin{array}{c c c c c c c c c}
|\nu_{bg}|      &               0       &       -|\nu_{bg}|     &       0               &       0                       &       0                       &       0               &       0       &       0               \\
        0       &       |\nu_{bg}|      &       0                       &       0               &       0                       &       0                       &       0               &       0       & -|\nu_{bg}    |\\
        0       &               0       &       |\nu_{bg}|              &       0               &       0                       &       0                       &       -|\nu'| &       0       &       0               \\
        0       &-|\nu_{bg}|    &       0                       &       |\nu_{bg}|      &       0                       &       0                       &       0               &       0       &       0               \\
        0       &               0       &       0                       &       0               & |\nu_{tip}|           &-|\nu_{tip}|+|\nu''|&  -|\nu''|        &       0       &       0               \\
        0       &               0       &       0                       &       0               & -|\nu_{tip}|+|\nu''|& |\nu_{tip}|     &       0               &-|\nu''|       &       0               \\
        0       &               0       &       0                       &       0               &       0                       &       -|\nu''|                &|\nu'|+|\nu''  |&      0       &       0               \\
        0       &               0       &       0                       &       -|\nu'| &       -|\nu''|                &       0                       &       0               &|\nu'|+|\nu''|&        0               \\
        0       &               0       &       0                       &-|\nu_{bg}|+|\nu'|&    0                       &       0                       &       0               &-|\nu'|        &       |\nu_{bg}|      \\
\end{array}\right)
\end{equation}
\end{widetext}
Given the matrix elements of the inverse $\boldsymbol{R}=\boldsymbol{G}^{-1}$, we can calculate the longitudinal resistance as
\begin{equation}
R_{xx} = R_{12,10\: 9} = \frac{-V_9}{I_1} = R_{92}-R_{91}.
\end{equation} 
This gives formula (\ref{eq:both_equilibration}) for $R_{xx}$.
For the special case $\nu'=\nu_{bg}$ shown in Fig.~\ref{fig:backscattering}(b), we obtain Eq.~(\ref{eq:indirect_equilibration}), and for the case  shown in Fig.~\ref{fig:backscattering}(a) with $\nu''=\nu_{tip}$, the formula (\ref{eq:direct_equilibration}). 

\begin{acknowledgments}
We thank Stefano Roddaro for useful discussions and Fabio Beltram for his continuous support. We acknowledge funding from the Italian Ministry of Foreign Affairs, Direzione Generale per la Promozione del Sistema Paese, and from the Polish Ministry of Science and Higher Education, Department of International Cooperation (agreement on scientific collaboration between Italy and Poland). Financial support from the CNR in the framework of the agreements on scientific collaborations between CNR and CNRS (France), NRF (Korea), and RFBR (Russia) is acknowledged. Funding from the European Union Seventh Framework Programme under Grant Agreement No. 696656 Graphene Core 1 is acknowledged. The theoretical side of this work was supported by the National Science Centre (NCN Poland)  according  to  decision  DEC-2015/17/B/ST3/01161, and by PL-Grid Infrastructure. S.~G.~acknowledges support by Fondazione Silvio Tronchetti Provera. S.~H.~acknowledges support from Scuola Normale Superiore, project SNS16\_B\_HEUN – 004155.
\end{acknowledgments}

\bibliography{SGM}

\end{document}